\documentclass{aa}

\usepackage{txfonts}
\usepackage{epsf}
\usepackage{graphicx}
\epsfclipon

\begin{document}
\title{Atmospheric dynamics in carbon-rich Miras}
\subtitle{II. Models meet observations}

\author{W.~Nowotny\inst{1} 
        \and T.~Lebzelter\inst{1}
        \and J.~Hron\inst{1}
        \and S.~H\"ofner\inst{2}}
\institute{Institut f\"ur Astronomie der Universit\"at Wien,
           T\"urkenschanzstra{\ss}e 17, A-1180 Wien, Austria 
      \and Departement of Astronomy and Space Physics, Uppsala University, 
           Box 515, SE-75120 Uppsala, Sweden
           }
\offprints{W. Nowotny,\\nowotny@astro.univie.ac.at}
\date{Received / Accepted}
\titlerunning{Atmospheric Dynamics in Carbon-rich Miras II.} 
\authorrunning{Nowotny et al.}

\abstract{
Originating in different depths of the very extended atmospheres of AGB stars, various molecular spectral lines observable in the near-infrared show diverse behaviours and can be used to probe atmospheric dynamics throughout the outer layers of these pulsating red giants. In Nowotny et al. (\cite{NAHGW05}, Paper\,I) time series of synthetic high-resolution spectra were presented, computed from a dynamic model atmosphere for a typical carbon-rich Mira. In this work, line profile shapes, their variations during the lightcycle and radial velocities derived from wavelength shifts are analyzed and compared with results from observed FTS spectra of the C-rich Mira \object{S Cep} and other Miras. It is found that the global velocity structure of the model is in qualitative agreement with observations. Radial velocities of molecular lines sampling different layers behave comparably, although some differences are apparant concerning absolute values. A correction factor of $p$\,$\approx$\,1.36 between measured RVs and actual gas velocities is derived for  CO $\Delta v$=3 lines. It is shown that dynamic model atmospheres are capable of reproducing Mira spectra without introducing an additional ''static layer'' proposed by several authors.

         }
\maketitle


\section{Introduction}   
\label{s:intro}

High-resolution (IR) spectroscopy has been successfully used to study the kinematics within all regions of the extended atmospheres of AGB stars, as summarised in detail in Nowotny et al. (\cite{NAHGW05}, in the following designated Paper \,I). The outer layers of these red giants are heavily influenced by pulsation of the stellar interior and by the development of a dust-driven stellar wind. From observations we know that different molecular lines sample layers of various depths and therefore show typical behaviours depending on the movement of gas at the corresponding optical depth.

Based on a dynamic atmospheric model, synthetic spectra were calculated by applying a radiative transfer that includes the influence of velocities. These were presented in Paper\,I, demonstrating that the models can qualitatively reproduce dynamics throughout the different regions of the atmosphere. Here, the results of our modelling (line profile shapes and their evolution as well as measured radial velocities) will be quantitatively compared to observations, especially to the ones of S\,Cep.


\section{Observational results of S\,Cep} \label{s:obsscep}

S\,Cep is the only carbon-rich Mira for which extensive time series IR spectroscopy has been obtained so far. As this paper deals with C-rich model atmospheres, S\,Cep was therefore chosen as the reference object to define the model parameters for our calculations. Observational results were published by Hinkle \& Barnbaum (\cite{HinkB96}, in the following HB96). S\,Cep is the brightest C-rich Mira in the northern sky and is believed not to be significantly obscured by dust (HB96).

HB96 obtained high-resolution spectra in the range of 1.6--2.5$\mu$m by using the FTS spectrograph at the 4m Mayall telescope at Kitt Peak. The spectral resolution was set to $R$$\approx$70\,000 and the S/N ratios are of the order of 50. Details can be found in their Table\,1. In addition to the published K--band radial velocities (RV), the spectra were kindly provided to us by K. Hinkle for the present comparison with our models. All spectra were corrected for earth motion and the stellar velocity\footnote{The center of mass radial velocity (CMRV) was derived from pure rotational molecular (CO, SiO, CS, HCN) emission lines in spectra observed with radio telescopes, see Table\,\ref{t:scepparameters}.} was subtracted.

Absorption lines of the CN molecule have been used in the past to study velocities in C star atmospheres (see references in Sect.\,2.3 of Paper\,I) as they are prominent in spectra of carbon stars. As the crowding of CN lines is very high, HB96 used cross--correlation of a region of 50\,cm$^{-1}$ width in the K--band to measure RVs, instead of line shifts of single lines. Also sampling deep photospheric layers, RV-curves from CN lines in the near IR (Fig.\,1 of HB96) show a similar behaviour to the ones for CO $\Delta v$=3 (Paper\,I). This was also found by Alvarez et al. (\cite{AJPGF01b}) from cross--correlation of visual spectra of S\,Cep (their Fig.\,3).

Figure\,\ref{f:rv-scep-beob} shows observed RVs taken from the literature (detailed explanations in HB96) as a function of visual phase $\phi_{\rm v}$. All measurements from different light cycles are combined and then repeated for better illustration. While CN $\Delta v$=--2 lines of the red system show the typical discontinuous RV-curve, CO $\Delta v$=2 low-excitation lines vary less and not regularly. The K\,I line, observable in the visual at $\approx$\,0.77$\mu$m, probes outflowing regions further out in the atmosphere (as CO $\Delta v$=1 lines) and shows RVs comparable to radio measurements of the circumstellar expansion (Table\,\ref{t:scepparameters}).

It is remarkable that the RV-curve of CN lines is asymmetric relative to the CMRV and extends to more negative velocities, while all observed CO $\Delta v$=3 RV-curves (e.g. Fig.1 of Lebzelter \& Hinkle \cite{LebzH02}) appear shifted to more positive values. 

\begin{table}[h]
\begin{center}
\caption{
Observed and derived properties of the carbon-rich Mira S\,Cep, taken from the literature:
1 -- from P-L-relation of Groenewegen \& Whitelock (\cite{GroeW96}),
2 -- $T_{\rm eff}$ and $m_{\rm bol}$ from Bergeat et al. (\cite{BerKR01}) + Hipparcos distance,
3 -- $m_{\rm bol}$ by interpolation/integration of $VJHKLM$ and IRAS magnitudes 
    (as described in Kerschbaum \& Hron \cite{KersH96}) + Hipparcos distance,
4 -- GCVS catalogue, Kholopov et al. (\cite{KhoSF88}),
5 -- Neugebauer \& Leighton (\cite{NeugL69}),
6 -- Whitelock et al. (\cite{WhiMF00}), 
7 -- Hipparcos catalogue, 
8 -- van Belle et al. (\cite{vBDTB97}), 
9 -- Eglitis \& Eglite (\cite{EgliE97}), 
10 -- Olofsson et al. (\cite{OlLNW98}), 
11 -- Bujarrabal et al. (\cite{BujGP89}), 
12 -- Loup et al. (\cite{LoFOP93}), 
13 -- Olofsson et al. (\cite{OlEGC93}),
14 -- Neri et al. (\cite{NKLBL98}),
15 -- Gautschy-Loidl et al. (\cite{GaHJH04}).}
\begin{tabular}{llcl}
\hline
\hline
&&&Ref.\\
\hline
$L$&[L$_{\odot}$]&7450&1\\
&&6792/8165&2\\
&&10814&3\\
\hline
$V$&[mag]&7.4\,--\,12.9&4\\
$I_0$&[mag]&5.25&5\\
$K_0$&[mag]&--0.15&5\\
&&--0.05&6\\
\hline
distance&[pc]&415$\pm$105&7\\
spectral type&&C\,7,4e (N\,8e)&4\\
$T_{\rm eff}$&[K]&2133$\pm$176&8\\
&&2095/2240&2\\
$R$&[R$_{\odot}$]&693$\pm$155&8\\
C/O&&1.194&9\\
&&1.4&15\\
\hline
$P$&[d]&486.84&4\\
$\Delta$V&[mag]&5.5&4\\
$\dot M$&[M$_{\odot}$\,yr$^{-1}$]&2$\cdot$10$^{-6}$&10,\,11,12\\
$v_{\rm exp}$&[km\,s$^{-1}$]&22&13,\,11,14\\
$CMRV_{\rm heliocent}$&[km\,s$^{-1}$]&--26.65&11,\,10,14\\
\hline
\end{tabular}
\label{t:scepparameters}
\end{center}
\end{table}


\section{Stellar parameters: real stars vs. models}
\label{s:parameters}

Table\,\ref{t:scepparameters} lists stellar parameters of S\,Cep derived from observations, which can be compared with the ones of the model atmosphere used in Table\,1 of Paper\,I. Relating these sets of parameters is not straightforward. In this respect, the dynamic atmospheric model used here has to be regarded as having justif\/iable and physically reasonable properties rather than being a dedicated "fit" to the C-rich Mira S\,Cep. Gautschy-Loidl~et~al. (\cite{GaHJH04}) were able to reproduce low-resolution spectra (ISO) rather well on the basis of this model and in this paper we further analyse the predictions of this model.

\subsection{Period, luminosity, amplitude, mass}  
\label{s:parameters1}

Only very few parameters are clearly derivable from observations and can directly be adopted for the modelling, like the pulsational period $P$ (490$^{\rm d}$ was used for the modelling). The luminosity $L$ can be constrained to some extent (Hipparcos distances). In this respect, the chosen mean luminosity of 10$^{4}$\,L$_{\odot}$ for the model is plausible (although probably at the high-luminosity end of the range). The amplitude $\Delta m_{\rm bol}$ of 0.86$^{\rm m}$ represents a realistic value comparable to observations of Mira variables (e.g. Whitelock et al. \cite{WhiMF00}). Only rough estimates can be given for stellar masses of galactic field stars such as S\,Cep. The chosen AGB mass of $M_*$=1\,M$_{\odot}$ is reasonable, although some statistical arguments (Barnbaum et al. \cite{BarKZ91}) point towards main sequence masses in the range of 2.5--4\,M$_{\odot}$ for certain C stars, including S\,Cep. In Sect.\,\ref{s:fitting} we also discuss a model with 1.5\,M$_{\odot}$.

\begin{figure}[t]
   \resizebox{\hsize}{!}{\includegraphics{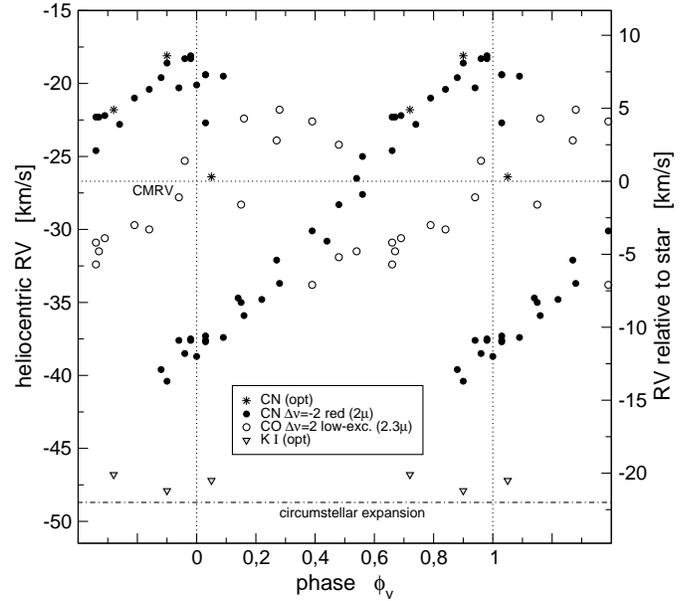}}
    \caption[]
    {Observed radial velocities derived from different lines in 
    high-resolution spectra of the carbon-rich Mira S\,Cep, compiled from HB96,
    Barnbaum (\cite{Barnb92b}) and Olofsson et al. (\cite{OlEGC93}).\\
    For S\,Cep: $v$=$v_{heliocent}$+26.65, $v_{heliocent}$=$v_{LSR}$--11.65
    [km\,s$^{-1}$]}
   \label{f:rv-scep-beob}
\end{figure}

\subsection{Effective temperature, surface gravity}  
\label{s:parameters2}

Relating the classical parameters $T_{\rm eff}$ and log~$g$ of observations to our models is hampered by two fundamental difficulties.

The first one is inherent to the dynamic models themselves. As described in detail in H\"ofner et al. (\cite{HoGAJ03}) the model calculation starts with a hydrostatic initial model. Being very similar to other model atmospheres for cool giants (as e.g. the ones computed with the MARCS code), they are characterised by the following fundamental parameters: luminosity $L_*$, mass $M_*$, effective temperature $T_*$ and the elemental abundances (the C/O ratio is especially important, all other abundances are assumed to be solar). These can be directly linked to the corresponding standard parameters of stellar atmospheres ($T_{\rm eff}$, $L$, log~$g$, chemical composition). Then stellar pulsation comes into operation, simulated by a variable inner boundary (piston), and dynamical  effects (extension, shock fronts, dust-driven winds) occur. The atmospheric structure is thus fundamentally different to the initial model at any point in time (compare Fig.\,9 in Paper\,I or Figs.\,3 and 4 in H\"ofner et al. \cite{HoGAJ03}). Not only is the spectral appearance thereby changed (Gautschy-Loidl \cite{Gauts01}), but also the structures can no longer be characterised by the parameters $T_*$ and $R_*$ (or log~$g$). In a more general context Baschek et al. (\cite{BasSW91}) assert that the classical term of "effective temperatures" is highly questionable for Mira stars. This topic was also discussed by Bessell et al. (\cite{BeBSW89a}).

The second problem is related to deriving values for $T_{\rm eff}$ and log~$g$ from observations of pulsating giants. The parameters available in the literature rely on interpreting observational data with the help of models (fitting synthetic spectra to observed ones). This process always implies uncertainties. This is especially so, as all studies in the past utilised hydrostatic model atmospheres. These may resemble AGB stars with only modest variability and mass loss quite well (e.g. J{\o}rgensen et al. \cite{JorHL00}), but obviously  are not an adequate approach for evolved red giants with more pronounced pulsations and strong outflows (as outlined e.g. in Hron et al. \cite{HAGHJ02} or Gautschy-Loidl et al. \cite{GaHJH04}).

Thus, in the end a direct comparison of $T_{\rm eff}$ and log~$g$, derived "inappropriately" from observations, with initial model parameters exhibiting a very indirect link to the actual dynamic model structure is somewhat misleading.

In the study presented here and in Paper\,I we use a dynamical model atmosphere possessing a $T_*$ of 2600\,K, which appears somewhat higher than the values for $T_{\rm eff}$ derived from observations of S\,Cep (Table\,\ref{t:scepparameters}). But -- this being only a parameter of the hydrostatic initial model -- it is not a colour temperature of the dynamic model and therefore not directly comparable. Despite some discrepancies, Gautschy-Loidl et al. (\cite{GaHJH04}) showed that synthetic spectra, based on the same model, resemble observed low-resolution spectra of S\,Cep in the IR rather well.

\subsection{Elemental abundances, C/O ratio}  \label{s:parameters3}

For C/O only the second problem discussed in Sect.\,\ref{s:parameters2} has to be considered. Gautschy-Loidl et al. (\cite{GaHJH04}) used the C$_3$ feature at 5.1\,$\mu$m as an indicator of C/O. The estimated value of C/O=1.4 (from the modelling and also from their comparison of observed spectra of several stars)  appears relatively plausible. Keeping in mind the limited accuracy (around 0.2--0.3) in determining this quantity, the ratios given in Table\,\ref{t:scepparameters} seem to be relatively consistent.

\subsection{Consequences}  \label{s:parameters4}

In spite of the difficulties in relating ($T_{\rm eff}$, log~$g$) to the fundamental parameters of dynamic models, the derivation of these parameters is definitely desirable. A major step in this direction has been made by Gautschy-Loidl et al. (\cite{GaHJH04}), who attempted to connect dynamical models and observations of real AGB stars for different phases by fitting spectra (e.g. ISO) and photometric indices. Starting from a grid of models they arrive at a "best fit model" for a given star by (f\/ine-)tuning the model parameters. Although they list the resulting fundamental parameters ($T_*$, $L_*$, $M_*$, C/O) of the corresponding hydrostatic initial model, Gautschy-Loidl et al. (\cite{GaHJH04}) emphasise that this "fitting" cannot (yet) be used to determine the stellar parameters for a given star. 

To proceed in this matter, further observational constraints have to be taken into account, for example detailed line profile studies (this paper) and intensity profiles/visibilities (e.g. Hron et al. \cite{HrNGH03}). Due to the complex structures of Mira atmospheres only such a multi-dimensional approach will lead to the desired results in the near future.


\section{Comparison of spectra and radial velocities} 
\label{s:complineprofiles}

We compare both line profiles and radial velocitiy variations in observed and synthetic spectra. The plotted observed spectra are normalised with respect to a nearby pseudo-continuum (for a discussion of FTS data reduction see Hinkle et al. \cite{HinWL95}). Synthetic spectra, on the other hand, are normalised relative to a computation with only the continuous opacity taken into account. In addition, a wavelength-independent value of opacity is added for the contribution of dust and molecules affecting the spectra pseudo-continuously (cf. Sect.\,5.1. in Paper\,I). A comparison of line depths of observations and calculations may therefore suffer from some uncertainties.

Synthetic radial velocity variations were derived from the shifts in wavelength of the resulting line profiles. This was done by: 
(i) direct measurements of the deepest point of the line, 
(ii) fitting splines through single profiles and measuring the minimum,
(iii) building the first moment of the flux distibution of the line or 
(iv) cross-correlation of a whole spectral region using the IRAF task \texttt{fxcor.} 
HB96 used a spectrum of BL\,Ori (C star with only small variations) as a template for the cross-correlation. Here, a spectrum for maximum light was used, computed without taking velocities into account. Synthetic velocities are then compared to observational results.

For the characterisation of radial velocities we followed the convention of observational papers, where velocities are defined as being negative for outflow (blue-shifted lines) and positive for infall (red-shifted lines).


\subsection{Visual and bolometric phases}
\label{s:phaseshift}

Observed RVs are usually linked to visual phases $\phi_{\rm v}$ within the lightcycle. For the model calculation we know the bolometric phases from the luminosity lightcurve (Fig.\,\ref{f:rv-scep-synth-absolut}). The question is how these are related. For O-rich Miras a small shift is suspected, such that bolometric phases lag behind visual ones. Lockwood \& Wing (\cite{LockW71}) found a typical shift of $\Delta\phi$$\approx$0.1 (if the phases measured by the filter at 1.04\,$\mu$m are taken to correspond to bolometric ones). Such a comparison has not been published for S\,Cep or other C-rich Miras yet. In Sect.\,\ref{s:rvs} a probable shift is discussed from a comparison of line profiles and RV-curves with observations. The shift is not expected to be very large, as the features dominating the $V$-band for C stars (CN, C$_2$) originate in deep photospheric layers with $T_{\rm gas}\approx$2500--4000\,K (Fig.\,3 of Loidl et al. \cite{LoHJA99}) and are less temperature sensitive than e.g. TiO in the O-rich case. Therefore the \mbox{$V$--lightcurve} should be closely coupled to the pulsation and the bolometric lightcurve. Also, lightcurves of C stars are more symmetric and possible phase shifts due to distortion should be minimal. An additional uncertainty may come from the fact that pulsation is simulated by a piston for the dynamic model atmospheres and does not come from a pulsation model (cf. H\"ofner et al. \cite{HoGAJ03}). In the following, \textit{bolometric phases} $\phi_{\rm bol}$ will be used to characterise \textit{synthetic} spectra and derived RVs, with numbers written in \textit{italic} for a clear distinction from visual phases $\phi_{\rm v}$ used to denote observations.


\subsection{Probing the pulsational layers}  \label{s:pulslayers}

\subsubsection{CO $\Delta v$=3 lines}  \label{s:co2ndovertone}

The second overtone CO lines were used most frequently to investigate deep photospheric layers in AGB stars and they show a very typical behaviour as already described in detail in Sect.\,5.4.1. of Paper\,I. Although some observations of these lines were reported by Barnbaum \& Hinkle (\cite{BarnH95}), identifications are difficult in C star spectra due to contamination by other molecules (CN, C$_2$). No time series studies have been published yet. The FTS spectra for S\,Cep of HB96 were all of poor quality in the H-band. Therefore, the synthetic line profiles in Fig.\,4 of Paper\,I can only be compared to observations of M- and S-type Miras. The behaviour also is supposed to be similar for C~stars. Fig.\,1 of Hinkle et al. (\cite{HinHR82}, in the following HHR82) shows averaged line profiles for different phases for the S-type Mira $\chi$\,Cyg, which will be used for the comparison here. A similar comparison was presented in Nowotny~et~al.~(\cite{NAHGH04}).

To match profiles of the same shape, a difference in phase of $\approx$0.25 between $\phi_{\rm v}$ of the $\chi$\,Cyg observations and $\phi_{\rm bol}$ of the model is needed. Line strengths are comparable and the synthetic profiles reproduce the typical pattern reasonably well, at least for the lower spectral resolution. However, the profiles do show some substructures at the spectral resolution of 300\,000 (Paper\,I). The shifts in wavelength are too small compared with observations and the doubling is less pronounced (compare phases $\phi_{\rm v}$=0.01/$\phi_{\rm bol}$=\textit{0.75}), but still clearly visible. This indicates that the difference in velocity behind and in front of the shockfront is too low in the atmospheric model (see Sect.\,\ref{s:largerampl}). A blue-shifted emission feature appearing before the line doubling phases (0.86/\textit{0.47}) can also be seen in the modelled profiles.

\begin{figure}
   \resizebox{\hsize}{!}{\includegraphics{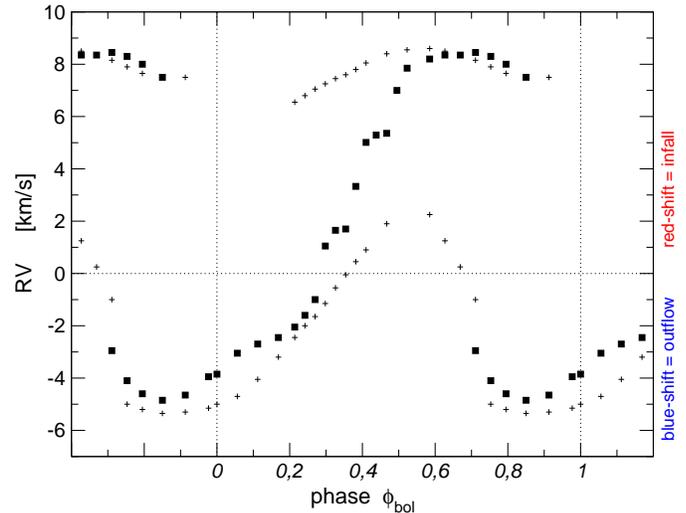}}
    \caption[]
    {Radial velocities for $\phi_{\rm bol}$=\textit{0.0--1.0}
    as derived from synthetic profiles of CO $\Delta v$=3 lines, 
    shown in Fig.\,4 of Paper\,I. Plotted for spectral
    resolutions of 300\,000 ({\tiny +}) and 70\,000 (squares) respectively.}
   \label{f:rv-scep-synth-CO16mue-phase}
\end{figure}

Figure\,\ref{f:rv-scep-synth-CO16mue-phase} shows RVs derived from these CO lines for various phases during one pulsational period. The developement of a few spectral components can be followed at the higher resolution of 300\,000. A weak component with $RV$$\approx$\,0\,km\,s$^{-1}$ seen at phases of \textit{0.6--0.7} may also (weakly) be recognised in the observed $\chi$\,Cyg profiles (phase 0.90 in Fig.\,1 of HHR82) as well as in the composite RV-curve of Lebzelter \& Hinkle (\cite{LebzH02}, Fig.\,1). It is conspicuous that line splitting is not only found around luminosity maximum, but also (but more moderately) during phases \textit{$\approx$0.2--0.5}, which is not being observed. This splitting turns into a continuous transition from blue- to red-shifts for a lower resolution of 70\,000, as the two components of the line profile melt into one broad feature (phase \textit{0.30}). Generally, the observed discontinuous RV-curve -- compare Fig.\,12 of HHR82 -- is reproduced by the low-resolution profiles. Only the RVs from the lower resolution were thus used in the following comparison (Figs.\,\ref{f:rv-scep-synth-absolut} and \ref{f:rv-scep-synth-phase}). The behaviour pattern is highly periodic (Fig.\,\ref{f:rv-scep-synth-phase}), as found in observations and as also is expected for the model (Paper\,I). Taking into account the above-mentioned shift of $\approx$\,0.25 between $\phi_{\rm v}$ and $\phi_{\rm bol}$, the "S-shape" is reproduced as well as the asymmetry\footnote{RV-curves extend to more positive values and appear "shifted" to $RV$=0\,km\,s$^{-1}$. This is seen in all observations of second overtone CO lines (e.g. Hinkle et al. \cite{HinSH84}, in the following HSH84). As one would expect to see equal infall and outflow for radial pulsations, this asymmetry is interpreted as due to line components that do not come from the same depth during the lightcycle. Note that also in the structures of the model atmosphere (Fig.\,9 in Paper\,I) infall velocities are larger than outflow in the pulsational layers.} w.r.t. $RV$=0\,km\,s$^{-1}$. Line doubling appears during phases \textit{$\approx$0.7--0.85}, a similar time interval as observed ($\Delta\phi$$\approx$\,0.2). Zero-crossing happens at $\phi_{\rm bol}$=\textit{0.3}, while it is observed at $\phi_{\rm v}$=0.4. Although the amplitude\footnote{maximum difference in $RV$ of the two components during phases of line doubling} $\Delta RV$$\approx$14\,km\,s$^{-1}$ is too low (typically 25\,km\,s$^{-1}$ in Mira observations), the characteristic behaviour could be reproduced successfully by consistent calculations.

\subsubsection{CN $\Delta v$=--2 lines}   \label{s:cnprofiles}

\begin{figure}[h]
   \resizebox{\hsize}{!}{\includegraphics{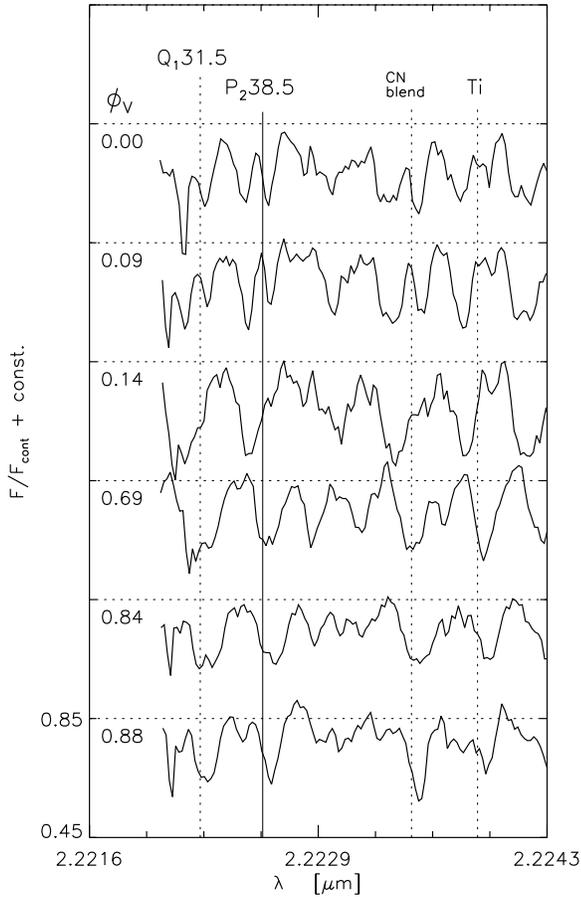}}
    \caption[]
    {Time series of high-resolution FTS spectra of S\,Cep (from HB96) 
    showing the 
    temporal behaviour of CN $\Delta v$=--2 lines of the red system. 
    The rest wavelengths taken from the literature are marked.}
   \label{f:cn2mueFTS}
\end{figure}

Figure\,\ref{f:cn2mueFTS} shows a sequence of FTS spectra of selected phases from the most thorough study of S\,Cep by HB96  with the relatively unblended 1--3 P$_2$38.5 line (most other regions of the K-band are much more crowded). It reveals that this line shows a similar behaviour as CO $\Delta v$=3 lines. For most of the CN lines there, this cannot be recognised clearly due to heavy blending. As it was noted by HB96 and was also found for R\,Leo by Hinkle \& Barnes (\cite{HinkB79b}), the IR lines of Ti (e.g. in Fig.\,\ref{f:cn2mueFTS} at 4496.625\,cm$^{-1}$) also sample pulsational layers and appear doubled around visual maximum, too.

\begin{figure}
   \resizebox{\hsize}{!}{\includegraphics{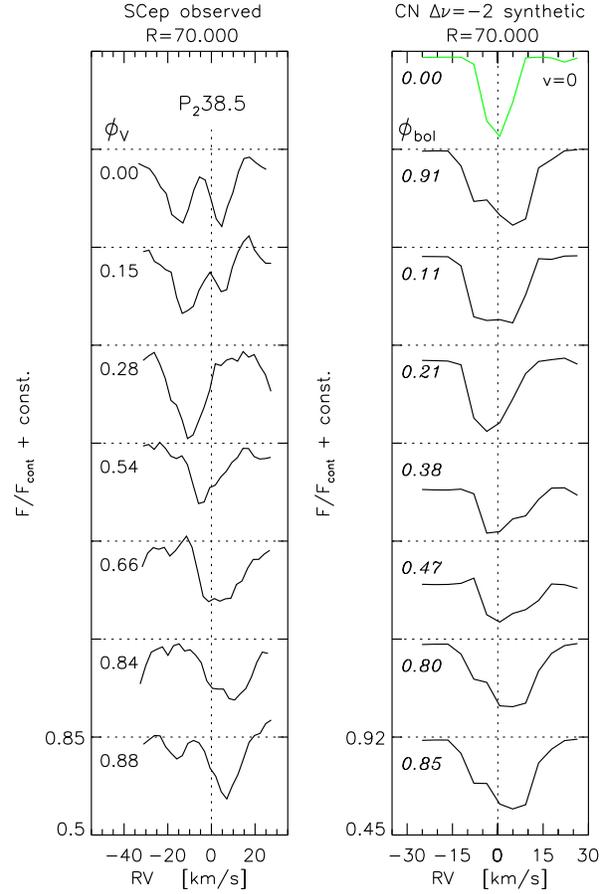}}
    \caption[]
    {Evolution of the observed CN 1--3 P$_2$38.5 line (taken from the FTS
    spectra
    of HB96) compared to synthetic line profiles for the CN 1--3 Q$_2$4.5 line;
    both can be found in the K-band around 2\,$\mu$m.}
   \label{f:cnprofiles}
\end{figure}

In Fig.\,\ref{f:cnprofiles} a comparison between the mentioned observed and synthetic spectral lines is presented. Although not exactly the same line -- observed CN 1--3 P$_2$38.5 vs. synthetic CN 1--3 Q$_2$4.5 -- the behaviour is expected to be very similar. Only the lower-resolution synthetic spectra from Fig.\,5 of Paper\,I are plotted here. The plotted profiles were chosen to show maximum congruence, which occurs not necessarily at the same phase/phase-shift. As already described in Paper\,I, the change with phase (red/blue-shifts, line doubling) can in principle be recognised from the synthetic profiles. While the ones at higher-resolution show more fine structures than known from observations, complexity is reduced by rebinning the spectra to lower resolutions and they become more similar to the observed ones. Line strengths are comparable, but again the shifts in wavelength are too small. The splitting of the two components around maximum phases is much too weak (0.00/\textit{0.91}). They merge and form one broad feature, while observations show two distinct lines. This is even more pronounced here than for CO $\Delta v$=3 lines in Sect.\,\ref{s:co2ndovertone}.

\begin{figure}
   \resizebox{\hsize}{!}{\includegraphics{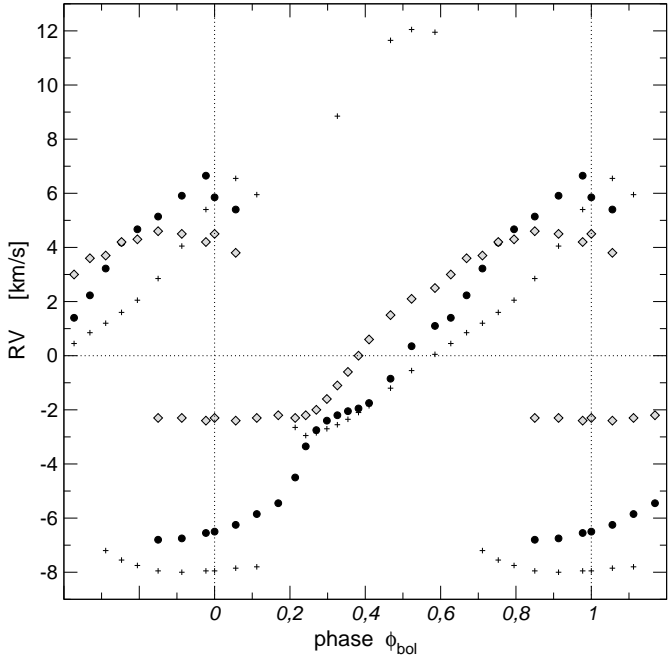}}
    \caption[]
    {Radial velocities for $\phi_{\rm bol}$=\textit{0.0--1.0}
    as derived from synthetic profiles of CN $\Delta v$=--2
    lines, shown in Fig.\,5 of Paper\,I. Measurements of single lines in spectra
    with resolutions of 300\,000 ({\tiny +}) and 70\,000 ($\bullet$) as well as
    results of the cross-correlation ({\tiny $\diamondsuit$}) are plotted.}
   \label{f:rv-scep-synth-CN2mue-phase}
\end{figure}

Figure\,\ref{f:rv-scep-synth-CN2mue-phase} shows RVs derived from synthetic CN lines during the first pulsational period examined. A plot of values measured from single profiles at highest spectral resolution looks even more complex than for CO lines due to the multi-component line profiles. Some components obviously present as a distortion of the line profile do not give a clearly measurable velocity. Opposite to the expectation (motivated by observed and R=70\,000 synthetic spectra) of one component (appearing at blue-shifts, moving toward red-shifts and disappearing again), the development of several components, which appear to be independent of each other, can be followed. But profiles rebinned to a lower resolution of 70\,000 lead to RV-curves ($\bullet$) that are in notable qualitative agreement with S\,Cep observations (Fig.\,\ref{f:rv-scep-beob}). Even the discontinuity in the blue-shifted component at $\phi_{\rm bol}$=\textit{0.15} is smoothed out.

\begin{figure}
   \resizebox{\hsize}{!}{\includegraphics{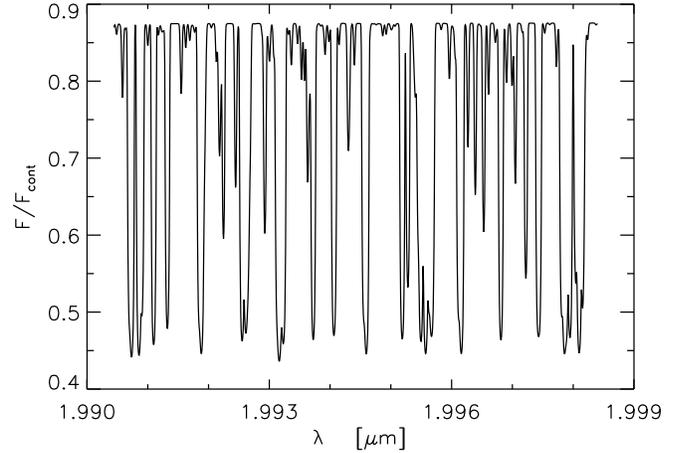}}
    \caption[]
    {Synthetic spectrum (resolution R=300\,000) dominated by CN 0--2 lines of 
    the red system
    (here for phase \textit{0.0} without velocities taken into account), 
    which was used as template for deriving radial velocities by 
    cross-correlation.}
   \label{f:cnspec}
\end{figure}

Another way to extract velocity curves is not to calculate one single line profile but rather a whole region occupied by CN lines and derive RVs by cross-correlation. This method is not able to resolve very weak components recognisable in single profiles as blended and unblended lines mix in the wavelength range used. HB96 used this technique because of the crowded CN spectra. We have tried to simulate this approach with synthetic spectra. Figure\,\ref{f:cnspec} shows the spectral region used, occupied by many CN $\Delta v$=--2 lines of the red system. It was chosen to have not too strong blending, but rather several strong individual lines. The positions of CN lines in the line list were not corrected for wavelength shifts between theoretical and true values; a direct comparison with any observed spectra may therefore not be straightforward. Relative velocities should still be correct. Also for this region spectra with resolutions of 300\,000 were calculated for all phases and the line shapes look exactly like the ones in Fig.\,\ref{f:cnprofiles}. RVs were then measured by cross-correlating these with the template spectrum shown in Fig.\,\ref{f:cnspec} ($\phi_{\rm bol}$=\textit{0.0}, no velocities taken into account in the radiative transfer). Results are also plotted in Fig.\,\ref{f:rv-scep-synth-CN2mue-phase}. The common characteristic of line doubling can be recognised and values similar to the ones from the low-resolution profiles are derived for phases $\phi_{\rm bol}$$\approx$\textit{0.25--0.8}. However, cross-correlation is not able to resolve the full amplitude in splitting for phases \textit{0.8--1.25}; the blue-shifted component ($RV$$\approx$--2.5\,km\,s$^{-1}$ instead of --7) is especially suppressed.

As they are more similar to observed ones, only RVs from low-resolution single profiles will be considered in the following and are plotted in Figs.\,\ref{f:rv-scep-synth-absolut} and \ref{f:rv-scep-synth-phase}, the latter demonstrating the highly periodic behaviour. The observed RV-curve of S\,Cep in Fig.\,\ref{f:rv-scep-beob} can be reproduced qualitatively by the synthetic ones. The scenario of line doubling due to shockfronts can be followed by CN line profiles, too. No global shift (as for CO $\Delta v$=3 lines) between observed $\phi_{\rm v}$ and $\phi_{\rm bol}$ of the models is needed to match the RV-curve of S\,Cep and the model. Line doubling appears almost for the same interval ($\phi_{\rm bol}$$\approx$\,\textit{0.85--1.1}). Zero-crossing occurs at the same phase (0.5). The synthetic RV-curve is symmetric w.r.t. $RV$=0\,km\,s$^{-1}$. The velocity amplitude of $\Delta RV$$\approx$13.5\,km\,s$^{-1}$ is too low compared with the observed one of 22\,km\,s$^{-1}$ for S\,Cep. Similar large amplitudes from CN lines in the optical have also been reported for a few other Miras by Barnbaum (\cite{Barnb92b}).


\subsection{Probing the dust-forming region -- CO $\Delta v$=2:}

\begin{figure}
   \resizebox{\hsize}{!}{\includegraphics{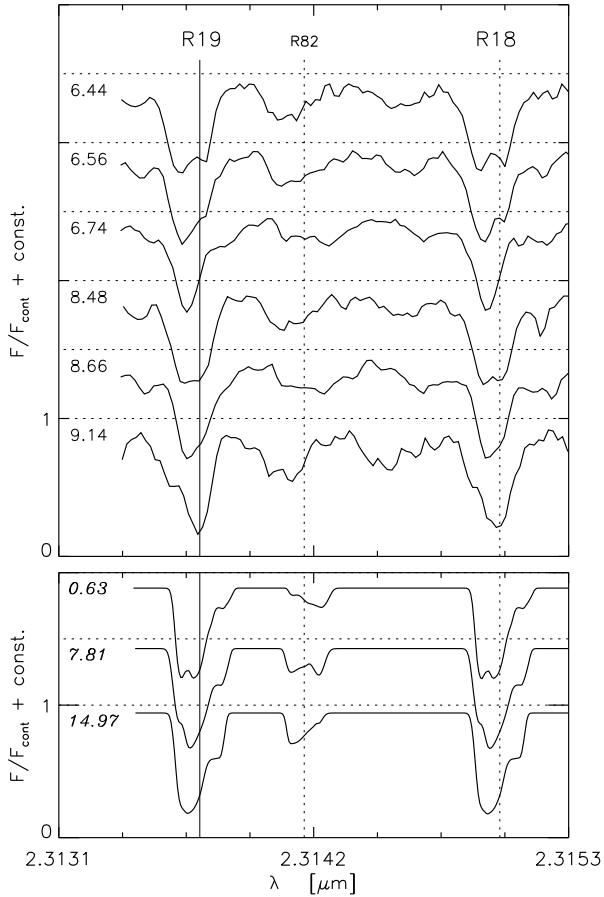}}
    \caption[]
    {\textit{Upper panel:} Observed CO $\Delta v$=2 lines from the FTS spectra 
    of S\,Cep (from HB96) for selected phases.\\
    \textit{Lower panel:} Synthetic line profiles ($R$=300\,000) for comparison.
    The 
    low-excitation lines (2--0 R19 and R18) consist of several components and 
    appear always blue-shifted, while the high-excitation line (2--0 R82) 
    follows the periodic behaviour of CO $\Delta v$=3 lines.}
   \label{f:co2mueFTSsynth}
\end{figure}

Figure\,\ref{f:co2mueFTSsynth} shows a comparison of first overtone CO lines of S\,Cep observations (HB96) with synthetic profiles, already presented and described in detail in Paper\,I. The weaker high-excitation lines (e.g. R82) exhibit the same behaviour as CO $\Delta v$=3 lines in the synthetic spectra, which can be difficult to see in observed C star spectra due to blending with several other lines. The very strong low-excitation lines (R18,19) can be identified clearly though and are studied in more detail here. While the profiles differ somewhat for different lines (compare e.g. R19 and R18) in spectra observed at a given time (due to contamination and observational uncertainties), they are exactly reproduced for synthetic ones. Line depths are comparable for synthetic and observed spectra as well as the general appearance (see also Fig.\,3 in HB96). These lines have complex shapes with various components present, although no clear line splitting is seen most of the time. Line profiles change with time, but these changes are not coupled to the lightcycle. This complex behaviour has also been found in obervations of R\,Leo by Hinkle (\cite{Hinkl78}) and of $\chi$\,Cyg by HHR82 (their Fig.\,8). Profile variations in the CO first overtone low-excitation lines that are not related to the light-cycle were also found earlier, e.g. by Winters et al. (\cite{WiKGS00}), who used a very similar approach.

Radial velocities were derived from the deepest points of the synthetic line profiles (at a resolution of 300\,000) and are shown in Figs.\,\ref{f:rv-scep-synth-absolut} and \ref{f:rv-scep-synth-phase} (open circles). Weak line doubling was found only in some cases, disappearing in rebinning to resolutions of 70\,000 (Fig.\,7 in Paper\,I). Sampling the region of the onset of the wind, the main component of low-excitation CO $\Delta v$=2 lines appears slightly blue-shifted by $\approx$5\,km\,s$^{-1}$ most of the time. In observed spectra, RVs for these lines were found to be either roughly at the CMRV ($\chi$\,Cyg) or show some variation (R\,Leo, S\,Cep or Lebzelter et al. \cite{LebHH99}), but with smaller amplitudes than lines sampling pulsational layers (CO $\Delta v$=3, CN, CO $\Delta v$=2 high-excitation). These three behaviours can be understood from the velocity structures of Fig.\,9 in Paper\,I, if the lines probe layers of 3, 2 and 1.5\,R$_*$ respectively. For all stars studied, RVs from CO $\Delta v$=2 lines were not coupled to the lightcycle. This may reflect the irregular motions within the dust-forming region (Sect.\,4 in Paper\,I), where CO $\Delta v$=2 lines originate.


\subsection{Probing the outflow -- CO $\Delta v$=1:}

As described in Paper\,I, lines of fundamental CO bands are usable to study the outflow of mass-losing Miras. Unfortunately, all of the few existing LM-band spectra for S\,Cep are of poor quality. A comparison can therefore only be done with other results. HHR82 presented in their Fig.\,11 a spectrum of $\chi$\,Cyg showing strong fundamental CO lines at 4.6$\mu$m. Being affected by several photosperic lines as well as telluric CO lines, they still show a constant blue-shift over time. The same lines were observed in spectra of IRC+10216 by Keady et al. (\cite{KeaHR88}) and later compared to models by Winters et al. (\cite{WiKGS00}). Profiles there compare well to the ones shown in Fig.\,8 of Paper\,I, except for the saturated absorption. They show P\,Cygni-type shapes, which is typical\footnote{This is however not necessarily the case for all lines being formed in the outflow. CO $\Delta v$=2 lines for example, although sampling the region where the wind starts, are not expected to show P\,Cygni-emission as the de-excitation $v$=2\,$\rightarrow$\,0 has a lower transition probability than cascading via the $v$=1-state.} for lines sampling the outermost regions and indicating stellar winds. RVs were calculated from the minima of the blue-shifted absorption in profiles with highest resolution (300\,000); the results are shown in Fig.\,\ref{f:rv-scep-synth-absolut}. The sometimes very broad or asymmetric absorption features can complicate the RV measurements. Apart from some small long-term variation in the terminal velocity, CO $\Delta v$=1 line velocities show a steady outflow over time.


\subsection{The overall picture}  \label{s:rvs}

\begin{figure*}
   \centering
   \includegraphics[width=17cm]{2572f08.eps}
    \caption[]
    {\textit{Upper panels:} Lightcurve of the dynamic model atmosphere
    for three different, separated periods. The phases 
    $\phi_{\rm bol}$ for which synthetic spectra were calculated and the
    reference phase \textit{0.00} are marked.\\
    \textit{Lower panels:} Compilation of the corresponding radial 
    velocities derived from 
    shifted line profiles for different types of molecular lines. 
    Note the velocity convention: Negative velocities (blue-shift) represent 
    outflowing matter, positive values (red-shift) denote infall.}
   \label{f:rv-scep-synth-absolut}
\end{figure*}

\begin{figure}
   \resizebox{\hsize}{!}{\includegraphics{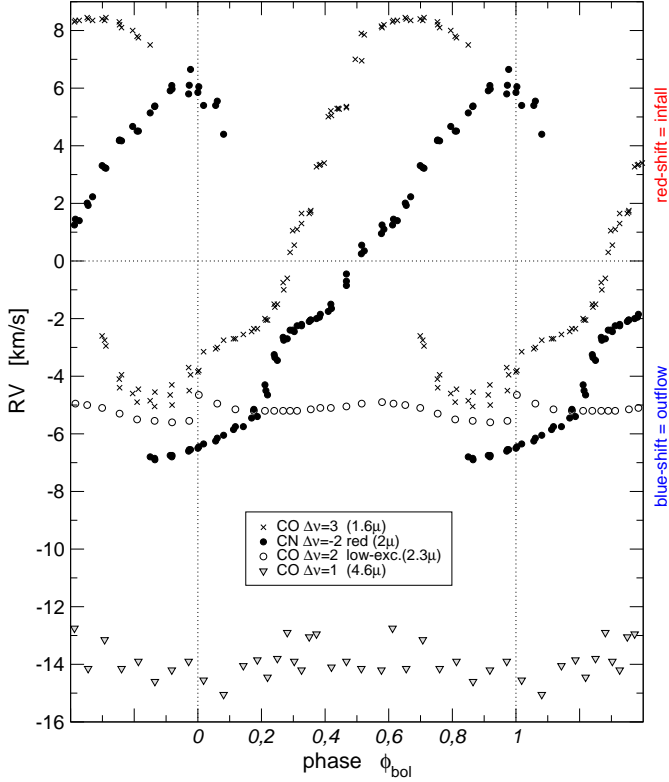}}
    \caption[]
    {Radial velocities from Fig.\,\ref{f:rv-scep-synth-absolut} projected onto
    one lightcycle. While CN and
    CO $\Delta v$=3 lines probe deep pulsating layers, CO $\Delta v$=2 lines
    come from the dust-forming region where the stellar wind is triggered
    and the outflow is sampled by CO $\Delta v$=1 lines.
    To be compared with observational results for S\,Cep in 
    Fig.\,\ref{f:rv-scep-beob}}
   \label{f:rv-scep-synth-phase}
\end{figure}                                         

Figure\,\ref{f:rv-scep-synth-absolut} shows a compilation of all RVs derived from synthetic line profiles presented above. Calculations were done for three separated periods (23 instances of time each). Figure\,\ref{f:rv-scep-synth-phase} shows a plot for direct comparison with the observations of S\,Cep in Fig.\,\ref{f:rv-scep-beob}. For CO $\Delta v$=3 and CN $\Delta v$=--2 lines  RVs from all periods were combined into one composite light-cycle and then plotted repeatedly for better illustration (RV-curves from different periods are almost identical each other, because of the very regular movement of the inner regions, see Fig.\,2 in Paper\,I). For CO $\Delta v$=2 and $\Delta v$=1 lines only measurements from one period ($\phi_{\rm bol}$=\textit{7--8} and \textit{14--15} respectively) are plotted for a clearer picture. It can be recognised that (with restrictions) our model calculations can reproduce the velocity pattern found for S\,Cep (and other Miras).

The first two types of lines sample deep pulsating layers and show typical discontinuous RV-curves, reflecting shockfronts running through the line-forming regions. Synthetic line profiles could generally reproduce this characteristic behaviour, at least for the lower spectral resolution of 70\,000 (only these results are adopted here). Some aspects (S-shape, asymmetry w.r.t. $RV$=0, line doubling interval) are realistically replicated with synthetic CO $\Delta v$=3 lines, but the apparant phase shift is conspicuous. While the mean phase of the line doubling interval is $\phi_{\rm v}$$\approx$1.0 in most observations (HSH84), it is clearly shifted to $\phi_{\rm bol}$$\approx$\textit{0.775} in Fig.\,\ref{f:rv-scep-synth-phase}, suggesting a shift of $\Delta\phi$$\approx$0.225 between visual and bolometric phases for our models, such that $\phi_{\rm bol}$ lags behind $\phi_{\rm v}$. Such a phase shift is not found for the RV-curve of CN lines however, which resembles the observations of S\,Cep in Fig.\,\ref{f:rv-scep-beob} rather well. Line doubling appears for the same interval and at the same phases, as well as zero-crossing. The synthetic RV-curve is symmetric around $RV$=0 (in contrast to CO lines), but the observed one even extends to more negative values. CN lines appear to originate slightly further out\footnote{Combining the difference of the median phases of line doubling for both type of lines ($\Delta\phi_{\rm bol}$$\approx$\textit{0.975--0.775}) with an estimation for the shock front propagation velocity from Fig.\,2 of Paper\,I (d$R$/d$\phi_{\rm bol}$$\approx$0.88 $\rightarrow u_{\rm front}$$\approx$7.2\,km\,s$^{-1}$) this would correspond to a difference $\Delta R$ of 0.176\,R$_{*}$/86.8\,R$_{\odot}$/0.4\,AU.} than the CO second overtone lines, which can be inferred from plots of the radial gradient of $\tau$ (as Fig.\,3 in Paper\,I) or from line doubling at later phases (the shockfront propagating outwards reaches outer layers later, also demonstrated in Fig.\,4 of Alvarez et al. \cite{AJPGF00}). Such a comparison would be interesting if done observationally in the future. Definitely, the amplitudes $\Delta RV$ of both types of lines are too small compared to observations (see Sect.\,\ref{s:largerampl}). 

Although synthetic profiles of CO $\Delta v$=2 low-excitation lines are comparable to observed ones, absolute values of RVs are qualitatively different. They show a small blue-shift at all times and therefore sample the region where the stellar wind starts ($\approx$3\,R$_*$). On the other hand, observed RVs are either constantly at $\approx$CMRV or vary little around the CMRV. Comparing Fig.\,9 of Paper\,I, this could be interpreted as either the lines should probe deeper layers ($\approx$2 or 1.5\,R$_*$) or pulsation should influence layers at larger radii.

Sampling the steady outflow, CO $\Delta v$=1 lines appear clearly blue-shifted, more or less constantly over time. Compared to observed RVs from K\,I lines in S\,Cep spectra (Fig.\,\ref{f:rv-scep-beob}), the synthetic ones are $\approx$5\,km\,s$^{-1}$ too low (as already expected from the velocity structures of the model, see Fig.\,9 in Paper\,I).

The velocity stratigraphy sketched for $\chi$\,Cyg around maximum light in  Wallerstein (\cite{Walle85}) can directly be recognised in the movement of mass shells of the model atmosphere for $\phi_{\rm bol}$$\approx$\textit{0.0}.


\subsection{Gas velocities vs. measured RVs}  \label{s:factorp}

To infer actual gas velocities from RVs of shifted spectral features, a correction factor \textit{p} -- with $u_{\rm gas}$=$p$$\cdot$$RV$ -- should be considered, caused by geometric aspects and center-to-limb effects. A first theoretical estimation of 24/17$\approx$1.41 for any radial pulsating star was given by Getting (\cite{Getti35}). Parsons (\cite{Parso72}) later found smaller values from synthetic lines calculated with Cepheid model atmospheres and a dependence on chosen parameters (his Fig.\,2). His approximate value of 1.31 was then used by Hinkle (\cite{Hinkl78}) to analyse the observations of R\,Leo. As the very extended Mira atmospheres have more complicated velocity structures (e.g. Fig.\,\ref{f:structure-scep1}) and lines from different wavelengths sample different depths, this correction becomes more difficult. As stated in Scholz\,\&\,Wood (\cite{SchoW00}) a factor of $\approx$1.4 was often used to interpret the observations. Comparing RVs derived from line shifts with gas velocities in front of and behind the shockfronts of their models, they derive factors of 1.25/1.4/1.6 for different lines (CO, OH) from model calculations. Using the same method, a correction factor of $p$\,$\approx$\,1.36 from synthetic CO $\Delta v$=3 lines at phases \textit{0.75/0.85} (Sect.\,\ref{s:co2ndovertone}) could be estimated here.


\section{Quasi-static, warm molecular envelopes and dynamic model atmospheres} \label{s:staticlayer}

Different velocity components have been found by studies of line profiles in observed high-resolution spectra of pulsating AGB stars in the past (HHR82). Among these, a layer with almost zero velocity relative to the stellar systemic velocity and an excitation temperature of about 800\,--\,1000\,K could be identified from low-excitation lines of CO ($\Delta v$=2) and OH. Being present in a number of Miras [e.g. in R\,Leo (Hinkle \cite{Hinkl78}) or $\chi$\,Cyg (HHR82)], this layer is variable in intensity. The variations do not follow the lightcycle and sometimes the static layer vanishes completely. HHR82 interpreted this layer as a reservoir for mass loss driven by radiation pressure on the dust, which can condense at these low temperatures.

Tsuji (\cite{Tsuji88}) found anomalies in low-excitation CO $\Delta v$=2 lines in high-resolution spectra of various M giants (with slightly higher temperatures and smaller light amplitudes). These could not be explained by photospheric absorption alone but seemed to be composed of two components. Subtracting synthetic spectra (based on hydrostatic model atmospheres) from observed ones revealed excess absorption with small blue- or red-shifts. Tsuji also interpreted this as being caused by a "quasi-static layer" in the outer atmosphere and derived excitation temperatures of 1000\,--\,2000\,K as well as mass estimates for it. Such a molecular formation zone could be well separated from the warm photosphere but not yet in the expanding regions of the cool stellar wind. Applying the same method to ISO SWS spectra, Tsuji et al. (\cite{TsOAY97}) found excess absorption/emission for other spectral features (molecular bands of H$_2$O, CO, SiO, CO$_2$), too. They claimed that such a "warm molecular layer" could be located at about 2\,R$_*$ and could be common not only in Miras, but in evolved red giants in general. Still it was not clear how such an extra component within a red giant atmosphere could develop and persist.

The atmospheric model used here shows a similar behaviour to that described above (compare Figs.\,2 and 9 in Paper\,I). Around 2\,R$_*$ we find a region where gas velocities are rather small at all times. In this transition zone between the pulsating photosphere and the cool outflow, dust is being formed (cf. Fig.\,1c in Paper\,I) and the stellar wind is triggered. Low-excitation CO $\Delta v$=2 lines, which sample these layers, show only small blue-shifts and RVs would be observable close to the stellar systemic velocity. This means that dynamic model atmospheres of H\"ofner et al. (\cite{HoGAJ03}) can produce a phenomenon resembling a quasi-static layer in a consistent way, and it is not necessary to introduce it artificially. This is at least true for some combinations of stellar parameters. As it will be shown in Sect.\,\ref{s:fitting}, a relatively small change in the chosen parameters (e.g. 1$\rightarrow$1.5\,M$_{\odot}$) can strongly influence velocities in the mentioned layers, due to the non-linear response of dust formation to the thermodynamic conditions in the upper atmosphere.

It will be interesting to perform a parameter study of various models with regard to this phenomenon. On the other hand observational evidence of a static layer for a statistically meaningful number of stars should be possible with the upcoming IR spectrographs. There is already some effort being spent to investigate existing high-resolution spectroscopic data (Lebzelter et al. \cite{LeNHH03}). In addition, interferometric measurements of Miras will help to tackle this question (e.g. Perrin et al. \cite{PRMCW04}).

The possible formation mechanism of "warm molecular layers" was investigated by Woitke et al. (\cite{WoHWJ99}) for O-rich Miras and by Helling \& Winters (\cite{HellW01a}, \cite{HellW01b}) for C-rich ones. They found enhanced molecular column densities in the layered structures of dynamic models of AGB outer atmospheres (compared with predictions of hydrostatic models), caused by stellar pulsation and propagating shock waves. Similar results regarding the partial pressures of selected molecules were described by Hron et al. (\cite{HLHJA98}) and Loidl et al. (\cite{LoHJA99}), also discussing the effects on the resulting synthetic spectra of C-rich models.


\section{Future intentions}   

\subsection{Fitting models to observations} \label{s:fitting}

As pointed out in Sect.\,\ref{s:rvs}, our model can qualitatively reproduce the velocities found for S\,Cep, but some differences remain concerning the absolute values. Therefore we tried to vary the input parameters of the model (Table\,1 of Paper\,I) to get even more realistic atmospheric structures.

\begin{figure}
   \resizebox{\hsize}{!}{\includegraphics{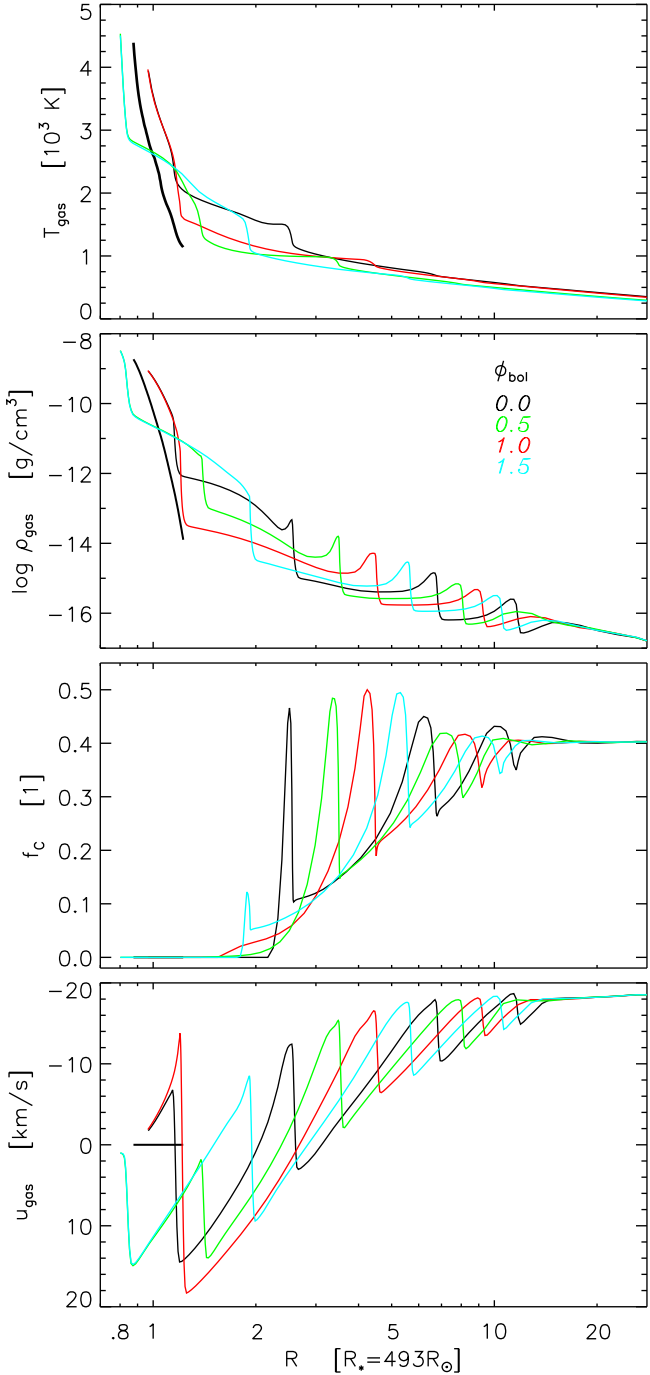}}
    \caption[]
    {Radial structures of a dynamic model atmosphere with parameters
    as in Table\,1 of Paper\,I, except for the stellar mass, which is
    changed to 1.5\,M$_{\odot}$. To be compared with structures of the original 
    model in Fig.\,1 of Paper\,I.}
    \label{f:structure-scep1}
\end{figure}

For example a model where only stellar mass is changed from 1 to 1.5\,M$_{\odot}$ was investigated. This seems comparably plausible, as we only have rough mass estimates for S\,Cep (e.g. 2.5--4\,M$_{\odot}$ from Barnbaum et al. \cite{BarKZ91}). Differences are immediately visible from the structures plotted in Fig.\,\ref{f:structure-scep1} (compare Figs.\,1 and 9 in Paper\,I). Dust shells and density enhancements are more strongly pronounced. The global velocity structure looks markedly different. The very regular motions due to pulsation only reach out to $\approx$1.3\,R$_*$. From there on the model structures are not necessarily similar for the same phases of different lightcycles any longer, because the dust formation cycle spans two pulsational periods. A plot with movements of mass shells (as Fig.\,2 in Paper\,I) looks similar to Fig.\,2b in H\"ofner et al. (\cite{HoGAJ03}). Throughout the dust-forming region strong variations in velocities can be seen; a steady outflow and a smooth distribution of the degree of dust condensation is not reached until $\approx$10\,R$_*$. 

These differences are also reflected in line profiles and RVs. While the RV-curve for CO $\Delta v$=3 lines is almost the same as for the original model, line strengths change considerably from one period to the next (duplicating only every second period). CN $\Delta v$=--2 lines are more affected by the double periodicity. Not only the line shapes (variable line strengths; line doubling much clearer), but also RV-curves (that are irregular with a larger $\Delta RV$ of $\approx$16\,km\,s$^{-1}$ for cross-correlation) evidently deviate. However, synthetic velocity amplitudes for lines from the inner regions could not be increased enough with this changed model to reproduce observed ones. While the faster outflow leads to more realistic RVs measured from CO $\Delta v$=1 lines ($\approx$18\,km\,s$^{-1}$), the largest differences by far compared to the original model are found for CO $\Delta v$=2 low-excitation lines. As suspected from the large variations of gas velocities within the dust-forming region at $\approx$2--3\,R$_*$ in Fig.\,\ref{f:structure-scep1}, these lines show various components, in part with even more pronounced variability than second overtone lines (which has never been observed so far).

From a few other studied models with slightly changed characteristics, it appears that the f\/ine-tuning to achieve a dedicated fit for the velocities of one particular star (like S\,Cep) is not a straightforward task (compare also the discussion in Sect.\,\ref{s:parameters}). Not only are the stellar parameters of S\,Cep poorly known ($L$, $M$, etc.), but a multidimensional grid of dynamic models would also be needed. Too many parameters interact and influence the velocity structures in subtle ways. Especially the transition region between regular pulsation and steady outflow (and thus the behaviour of first overtone low-excitation CO lines) seems to depend sensitively on the chosen set of stellar parameters. This dependency was discussed also by e.g. H\"ofner \& Dorfi (\cite{HoefD97}) or Winters et al. (\cite{WLJHS00}).

\subsection{Larger velocity amplitudes} \label{s:largerampl}

The initial aim of future calculations will be a dynamic model atmosphere with a more realistic velocity amplitude within the pulsational layers. As stated in Sect.\,2.2 of Paper\,I, the RV-curves of CO $\Delta v$=3 lines show a rather uniform picture for all Miras studied so far, independent of their properties. Figure\,1 of Lebzelter \& Hinkle (\cite{LebzH02}) and the results in HSH84 demonstrate that amplitudes of $\Delta RV$$\approx$20--30\,km\,s$^{-1}$ can be observed. This common feature seems to be a fundamental characteristic of Miras and should be reproduced by a realistic model. The $\Delta RV$ values given in Scholz \& Wood (\cite{SchoW00}) could serve as a guideline for subsequent modelling. They state that a typical Mira appears to have a true (observed RVs corrected by their factor \textit{p} to get real gas velocities, Sect.\,\ref{s:factorp}) post-shock outflow velocity of $\approx$--14\,km\,s$^{-1}$ and a pre-shock infall velocity of $\approx$20\,km\,s$^{-1}$, resulting in a full velocity amplitude of $\approx$34\,km\,s$^{-1}$. Comparing this with gas velocities in the inner regions of the model used here (Fig.\,9 in Paper\,I), it can be deduced that the dynamic model should have an amplitude larger by a factor of $\approx$2. With these more extreme velocity gradients across the shockfront, line doubling should be more pronounced for CO $\Delta v$=3 and CN lines and easier to recognise even in low-resolution spectra (Sect.\,\ref{s:pulslayers}). From our studies so far we know that a larger shock amplitude cannot be achieved simply by applying a piston with a higher velocity amplitude $\Delta u_{\rm p}$. This is due to the self-regulating interdependence of the density structure (levitation) and the maximum shock strength. More detailed work is needed to address this problem.

\subsection{O-rich model atmospheres}

Calculations presented here and in Paper\,I focus on carbon-rich model atmospheres, as the formation and evolution of C-rich dust is much better understood and implemented in these models. More data have been obtained observationally for M-type stars. Thus further investigations of synthetic line profiles (for CO lines but also for others found in spectra of O-rich stars, as OH, SiO, etc.) based on oxygen-rich models may be revealing.


\section{Conclusions}

From high-resolution spectroscopic studies it is known that different molecular absorption lines in the NIR sample different depths of the dynamic atmospheres of pulsating AGB stars. Gas velocities there can heavily influence line profiles (Doppler shift, broadening, line doubling). Reproducing this scenario by consistent calculations is an indicator of the quality of dynamic model atmospheres. The advanced models of H\"ofner et al. (\cite{HoGAJ03}) allow us to simulate the dynamics in C-rich Mira stars from the innermost layers dominated by pulsation out to the regions of the stellar wind. Synthetic line profiles (presented in Paper\,I) were compared in detail with observations of S\,Cep and other Miras. In addition, radial velocities were derived from their shifts in wavelength.

It was found that the characteristic behaviour of spectral features sampling different regions (pulsational layers, dust-forming region, outflowing outer atmosphere) could qualitatively be reproduced by consistent calculations. Line profiles compare reasonably well to observed ones. Photospheric lines of CO ($\Delta v$=3) and CN \mbox{($\Delta v$=--2)} appear more complex at highest spectral resolution. Rebinning to lower resolutions (comparable to observed FTS spectra) leads to the expected, very typical S-shaped RV-curve. Together with RVs of low excitation CO $\Delta v$=2 lines (sampling layers where the stellar wind is triggered) and of CO $\Delta v$=1 lines (probing the outermost regions with strong outflow) we made a direct comparison with the equivalent data for S\,Cep. This comparison reveals that the dynamic model atmosphere used can reproduce -- at least qualitatively -- the global velocity structure throughout the atmosphere of a C-rich Mira.

Nevertheless, there are still quantitative differences: for example both the RV amplitudes in the inner regions and the outflow velocities are too small. Generating a model with quantitatively fitting radial velocities for a  particular observed star will be challenging (Sect.\,\ref{s:fitting}). A  realistic model should have a larger velocity amplitude in the pulsating photosphere, which appears to be a common feature of Mira stars (Sect.\,\ref{s:largerampl}).

The interplay between pulsation and the onset of a dust-driven wind can result in a transition zone with rather low gas velocities, at least for models with certain parameters. Thus, the dynamic model atmospheres used can provide a natural and phyically consistent explanation for a supposed quasi-static layer within the atmosphere, which was proposed for some red giants from spectroscopic observations (Sect.\,\ref{s:staticlayer}).

In addition to high-resolution spectroscopy, interferometric (narrow band) observations with high spatial resolution (e.g. Hron et al. \cite{HrNGH03}) will be a complementary major tool for studies of Mira atmospheres in the future. Time series of synchronous observations with both methods could provide unprecedented insights and strong constraints for atmospheric modelling.


\begin{acknowledgements}
We wish to thank Ken Hinkle for providing FTS spectra of S\,Cep. Sincere thanks are given to B. Aringer for careful reading and fruitful discussions. This work was supported by the ``Fonds zur F\"orderung der Wis\-sen\-schaft\-li\-chen For\-schung'' under project number P14365--PHY and the Swedish Research Council. TL receives an APART grant from the Austrian Academy of Science. We used the Simbad database operated at CDS, Strasbourg, France.

\end{acknowledgements}

\end{document}